\documentclass[final
    ,final            
  ]
  {aipproc}

\layoutstyle{8x11single}

\begin{document}

\title{Using evidence to make decisions}


\author{Charles Jenkins}{
  address={CSIRO, Pye Laboratory, Black Mountain, Canberra\\ \tt{charles.jenkins@csiro.au}}
 }

\begin{abstract}
Bayesian evidence ratios give a very attractive way of comparing models, and being able to quote the odds on a particular model seems a very clear motivation for making a choice.  Jeffreys' scale of evidence is often used in the interpretation of evidence ratios.  A natural question is, how often will you get it right when you choose on the basis of some threshold value of the evidence ratio? The evidence ratio will be different in different realizations of the data, and its utility can be examined in a Neyman-Pearson like way to see what the trade-offs are between statistical power (the chance of ``getting it right'') versus the false alarm rate, picking the alternative hypothesis when the null is actually true.  I will show some simple examples which show that there can be a surprisingly large range for an evidence ratio under different realizations of the data. It seems best not to  simply rely on Jeffrey's scale when decisions have to be taken,  but also to examine the probability of taking the ``wrong'' decision if some evidence ratio is taken to be decisive. Interestingly, Turing knew this and applied it during WWII, although (like much else) he did not publish it.
\end{abstract}

\maketitle

\section{Introduction}
The evidence, in a Bayesian context, is the denominator in Bayes' formula.  Its virtues do not need advocacy in this volume: being able to quote the odds, in favour of a hypothesis in the light of the data, is an extremely attractive feature of the Bayesian method.  The related quantity, the Bayes factor (here ${\cal B}$), multiplies the prior odds  on two opposing hypotheses to give the posterior odds.  The log (any base) of the Bayes factor is called the ``weight of evidence''.  All this is explained in many places; the book by Lee
\cite{lee04} is a good lubricant for rusty memory.

Jeffreys was the originator of a handy classification of values of the weight of evidence, which is often reproduced.  In Table \ref{tab:a} is a slightly amended version of Jeffreys' original, from Kass and Raftery \cite{kr95}. This sort of table appears to tell us what to do with our calculations, but  two questions arise.  Firstly, ``decisive'' (for example) -- in what circumstances?  Deciding to bet a dollar, or commit the D-Day landing forces, are not the same sort of decision because the consequences are so different. Secondly, the weight of evidence is just another statistic; if we repeat our experiment, we may get different data, and a different value for the weight of evidence.  This fact is implicit in the very definition of ${\cal B}$, which involves the likelihood of the data and hence its probability distribution. If the weight of evidence is variable, we may be interested in its sampling distribution because we will want to know the answer to questions like, ``if the null hypothesis holds, how often will I still get decisive evidence against it?''.To be fair to the weight of evidence, it only tells us about the weight of evidence \emph{contributed by the data}, and so naturally it does not take into account context or prior odds.  However it will be interesting to see how the categories fit in to the end-to-end process of taking a decision.

\begin{table}[!hbp]
\begin{tabular}{rrr}
\hline
   \tablehead{1}{r}{b}{$\log_{10} {\cal B}$\\}
  & \tablehead{1}{r}{b}{$ {\cal B}$\\}
  & \tablehead{1}{r}{b}{Evidence against \\the null hypothesis}
   \\
\hline
0 to 1/2 & 1 to 3.2 & Not worth more \\& & than a bare mention \\
1/2 to 1 & 3.2 to 10 & Substantial \\
1 to 2 & 10 to 100 & Strong \\
>2 & >100 & Decisive \\
\hline
\end{tabular}
\caption{Standard categories of the weight of evidence}
\label{tab:a}
\end{table}

\section{Variability in the Bayes factor}

\subsection{A spectroscopic example}

\begin{figure}
  \includegraphics[height=.3\textheight]{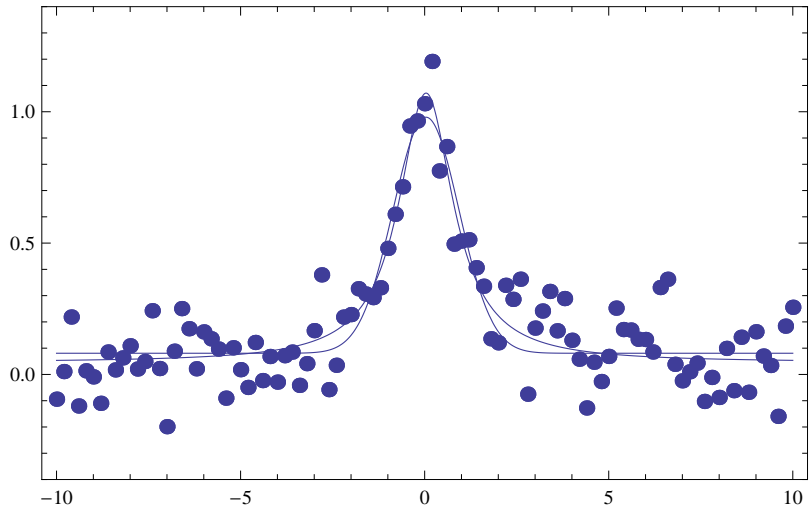}
  \caption{A simulated spectrum, dots, containing Gaussian noise, and best-fits of Gaussian and Lorentzian line profiles.}
  \label{figure1}
\end{figure}

Here (Figure \ref{figure1}) is a simple example to illustrate the issues.  It is a simulation of some spectroscopic data, a single line, of quite low signal-to-noise ratio (only 10 at the peak).  
I then calculate the Bayes factor given two hypotheses, which are that the line profile is either Gaussian or Lorentzian. I take these hypotheses, at first in this discussion, to be equally probable a priori.  It turns out that the posterior probability distribution for this little problem is close to a Gaussian, so I use the Laplace approximation to compute the Bayes factor.  This approximation of course does not hold in general but is a computational detail that does not affect the results. In the Figure I show the best fits (maximum of the posterior) of the two models, to illustrate what is apparent to any experimentalist - these two line profiles will be hard to distinguish from each other unless the signal-to-noise ratio is rather good.  However, for the example plotted, the Bayes factor gives the odds in favour of  any Lorentzian profile as 335:1  This seems surprising, given the quality of the data, but of course these odds represent an average over all possible fits, and the appearance of the best-fit in Figure \ref{figure1} may mislead. This kind of result prompted me to re-run the simulation with other realizations of the data, and it soon became apparent that there was variation in the Bayes factor. Now clearly the Bayes factor must vary if the input data are varied, but I found the \emph{amount} of variation  quite surprising, as I will now discuss.

\begin{figure}
  \includegraphics[height=.3\textheight]{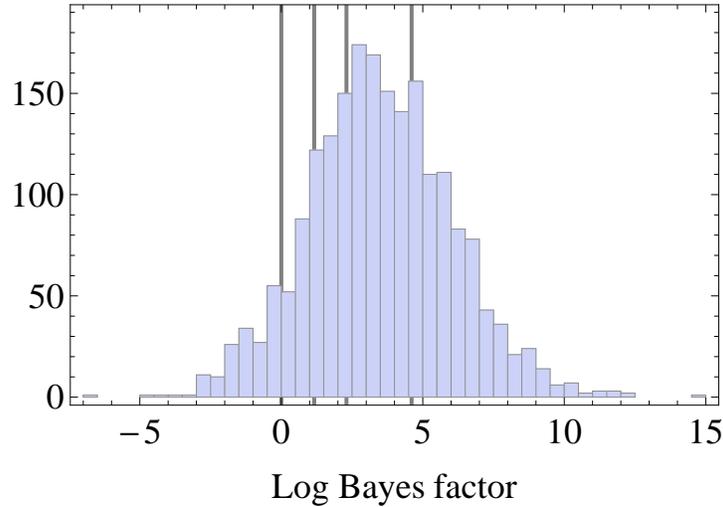}
  \caption{A histogram of the weight of evidence, using the natural logarithm (hence the units are ``natural bans'').  The histogram uses the results of 1000 repetitions of the line-fitting simulation. Vertical lines mark the boundaries of the categories of Table 1; from left to right, only worth a mention, substantial, strong, and decisive.}
  \label{figure2}
\end{figure}

Repeating this experiment many times gives a histogram of the weight of evidence, as shown in Figure \ref{figure2}.  The various categories are also shown, and we see that the possible values are spread through all the categories, from ``not worth a mention'' to ``decisive''.  The mean of the weight of evidence corresponds to odds of around 33 to 1, which seems too good to be true to an experimentalist's eye: this is probably because experimentalists know that the tiny differences in the wings of the profile would not be reliable in the real world.  In fact, adding a small amount of extra sophistication to the models, in the form of a possible low-order polynomial baseline, brings these odds down by an order of magnitude. Also, the noise is often correlated in real data.  But these are  separate issues.

\subsection{Generality}

Is it generally true that the statistical spread in the weight of evidence is comparable to its mean, or is this just a peculiarity of a particular numerical example?  I think it is general, for a variety of heuristic reasons.  I am not aware of any general proofs.

Jenkins and Peacock \cite{jp11} examined several simple cases and showed that this type of behaviour can be found analytically. In particular they showed that the mean and variance of the weight of evidence were comparable.


A heuristic argument is as follows.  If the Laplace approximation holds (so that the posterior distribution is close to a multivariate Gaussian) then the Bayes factor is  a ratio of maximum likelihoods, multiplied by a ratio of factors that depend on the width of the posterior (this term is related to the ``Ockham Factor'' and penalizes extra parameters in the models).  A well-known theorem \cite{mgb} tells us that the logarithm of a likelihood ratio (in certain, not too restrictive circumstances) will be distributed like chi-square.  The variance of a chi-square distribution is twice its mean, so this may be  telling us that the weight of evidence will have considerable spread about its mean. The argument is incomplete however because I have not dealt with the possible variability in the Ockham factor term, but I have not found a way to complete the proof.

Finally, Turing knew that the weight of evidence was very variable, which certainly gives the observation a good pedigree.  In his memoir on Turing's unpublished war work, Good \cite{good79} says \begin{quote}
Turing considered a model in which the weight of evidence $W$ in favour of the true hypothesis $H$ had a normal distribution, say with mean $\mu$ and variance $\sigma^2$. He found, under this assumption, (i) that if $H$ is false $W$ must again have a normal distribution with mean $-\mu$ and variance $\sigma^2$ and (ii) that $\sigma^2=2\mu$ when natural bans were used.
  \end{quote}
Here natural bans means that the logarithm in the weight of evidence is the natural logarithm.  Turing's result is the same as my conclusion based on the likelihood ratio argument, but I have not been able to find how he proved it.  It is possible that he used the likelihood ratio as well, and  the fact that a chi-square distribution is asymptotically Gaussian.

While none of the above points are proofs, they certainly seem to indicate that one should check the variance before making weighty decisions on the basis of a Bayes factor.

\section{Taking decisions}
\subsection{The Neyman-Pearson approach}

In the Neyman-Pearson approach, we focus on questions like ``If the line profile is indeed Lorentzian ($H_1$, what is the chance that I will get ${\cal B}>{\cal B}_{\rm{thresh}}$?'' (the \emph{power} ${\cal P}$) and ``If the line profile is not Lorentzian but Gaussian ($H_0$), what is the chance that I will still get ${\cal B}>{\cal B}_{\rm{thresh}}$?'' (the \emph{false alarm rate} $\alpha$).

\begin{eqnarray}
{\cal P} & = & \rm{prob}({\cal B}\ge {\cal B}_{\rm{thresh}} | H_1) \\
\alpha & = & \rm{prob}({\cal B}\ge {\cal B}_{\rm{thresh}} | H_0) \\ \nonumber
\end{eqnarray}

The idea is that you decide in advance what threshold in evidence you need to perform a certain action, and if the threshold is exceeded, you take that action.  The ideas are classical rather than Bayesian in form, but have obvious utility.  A commander may for example be evaluating a radar identification and targeting algorithm; the power helps with the question, ``If there are enemy aircraft in my airspace, what is the chance that this algorithm will direct me to fire at them?''

\begin{figure}
  \includegraphics[height=.3\textheight]{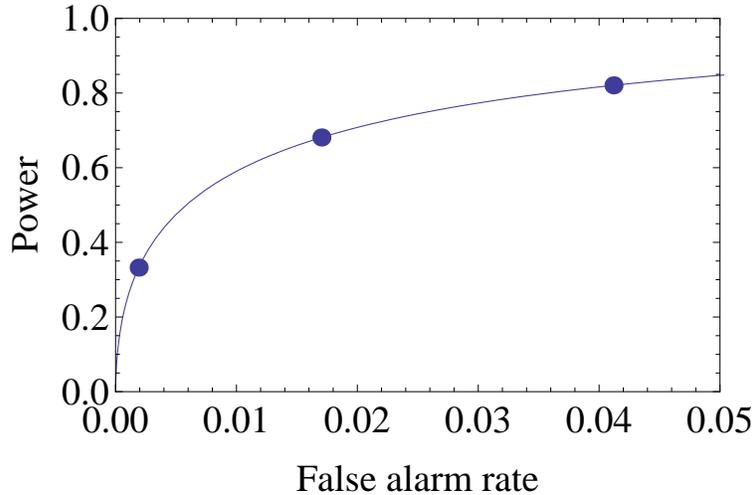}
  \caption{The power is plotted against the false alarm rate, for the Lorentzian vs. Gaussian profile fitting example.  The parameter that varies along the line is the threshold in weight of evidence, with ``decisive'' values to the left of the leftmost circle, ``strong'' between that circle and the next to the right, and ``substantial'' to the right of the rightmost circle.}
  \label{figure3}
\end{figure}

Given results such as those of Figure \ref{figure2}, we can produce the ``Receiver Operating Characteristic'' diagram, plotting power against false alarm rate -- Figure \ref{figure3}.  The name of the diagram presumably reflects its origins in radar research.
There is always a trade-off between power and false alarm rate, as shown here.

We see that, viewed in this way, the Jeffreys criteria are quite conservative.  Requiring decisive evidence before acting (in our case, deciding that the line profile really is Lorentzian) means that the power is less than 40\% but the false alarm rate is correspondingly small.  For the power to be high, we have to be relaxed about false alarms.  In the military case, a false alarm might mean shooting down one of your own pilots, so values are inherent in the use of these diagrams.  I will return to the notion of value later.

\subsection{Positive Predictive Power}
\begin{figure}
  \includegraphics[height=.3\textheight]{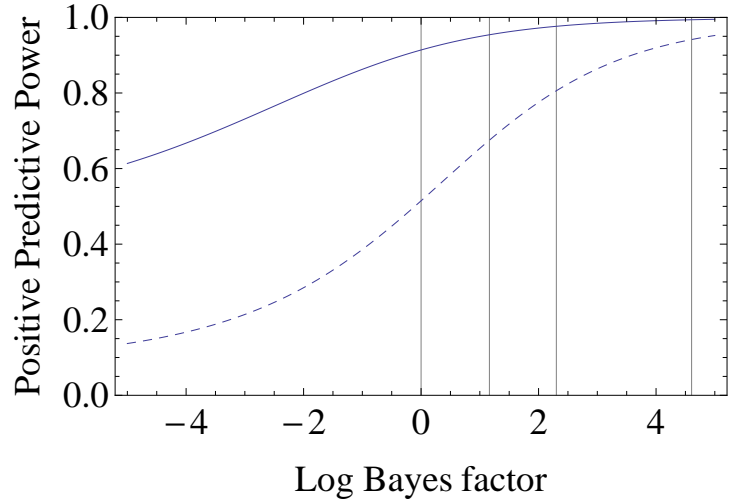}
  \caption{The PPP is plotted against the weight of evidence threshold, with the boundary of the Jeffreys categories illustrated from left to right by vertical lines; ``only worth a mention'', ``substantial'', ``strong'', and ``decisive''.  The solid line corresponds to $\Theta=1$ and the dashed line to $\Theta=1/10$.}
  \label{figure4}
\end{figure}

The Positive Predictive Power (PPP) is the Bayesian image of the Neyman-Pearson power:  in terms of our example, the PPP is the probability that the line profile is indeed Lorentzian, given that we have obtained a value for the weight of evidence above some pre-determined threshold. Knowing the power ${\cal P}$ and the false alarm rate $\alpha$ at some threshold value of the weight of evidence, the PPP follows from Bayes' Theorem:

\begin{equation}
{\rm PPP}=\frac{\Theta {\cal P}}{\alpha+ \Theta {\cal P}}
\label{eqn1}
\end{equation}
in which the quantity $\Theta$ is the prior odds ratio.  In our example, it is the prior odds on the Lorentzian hypothesis. In Figure \ref{figure4} the PPP is plotted for a couple of cases, together with the usual lines marking the boundaries between Jeffrey's categories. Interestingly, if the prior odds ratio is unity, we see that the PPP is very high for any category; ``barely worth a mention'' weighs in at over 90\%. Why is this?  Looking at Equation \ref{eqn1}, we see that it is because the false alarm rates, for all the Jeffreys categories, are so low.  This was the point about conservatism made in connection with the Neyman-Pearson approach.

In real life we are often sceptical about the ``alternative hypothesis''.  If we take the prior odds ratio in favour of the upstart Lorentzian hypothesis to be only 1/10, we get the dashed line in Figure \ref{figure4}, where we see that the Jeffreys categories make more useful distinctions. This shows clearly how the significance attached to the weight of evidence is context-dependent.  There is a sense in which being risk-averse (wanting a very low false alarm rate, or being \emph{a priori} sceptical) seems to make sense of the Jeffreys categories. This is because the weight of evidence is about the contribution of the \emph{data} to our decision-making, and the data speak more clearly if the result is not a foregone conclusion on other grounds.

It is worth noting that the PPP is useful because it is not restricted to two hypotheses, a null and an alternative, as is the case with Neyman-Pearson.

\subsection{Utilities and decisions}
\begin{figure}
  \includegraphics[height=.3\textheight]{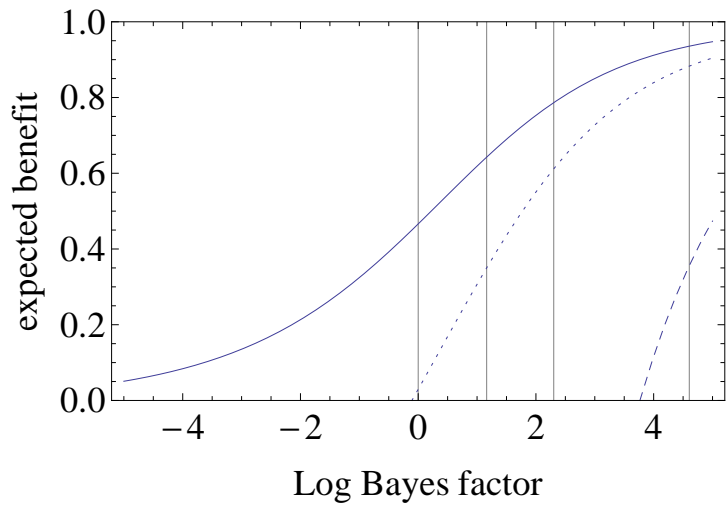}
  \caption{The normalized expected benefit, as defined in Equation \ref{eqn2}, is plotted against the weight of evidence threshold; from left to right, vertical lines mark the boundaries of the categories ``only worth a mention'', ``substantial'', ``strong'', and ``decisive''.  The solid line is for |loss/gain|=1/10, the dotted line for |loss/gain|=1 and the dashed line for |loss/gain|=10. }
  \label{figure5}
\end{figure}

There is only a decision if there are upsides and downsides.  Suppose our scientific life is so exciting that there is an actual immediate payoff from our decision about the shape of the line profile.  We can write the normalized expected utility as
\begin{equation}
B = PPP - |\frac{ \rm{loss} }{\rm{gain}} | (1-PPP),
\label{eqn2}
\end{equation}
assuming the loss is a negative number.  In Figure \ref{figure5} the expected benefit is plotted for several values of loss/gain, taking the prior odds $\Theta$ to be the sceptical 1/10.  Again we see that if there is a daunting downside (|loss/gain|=10) then the Jeffreys categories make sense; but are less discriminatory if the upside dominates.

\section{Conclusions}
The interpretation of the Bayesian weight of evidence, in terms of its ability to support decisions, has been examined.  It seems clear that the weight of evidence is quite noisy upon repeated trials, with its variance about equal to its mean.  This has implications for its use in practice, because a ``decisive'' value may be a statistical outlier and lead to the wrong decision.  Standard tools from decision theory can be used to understand the implications, but any decision of importance will involve value judgements. The systematic study of these values is the important field of utility theory, much discussed in economics. 

Three positive points to close.  One is that the mean of the weight of evidence improves very rapidly with the signal-to-noise of the data, so if that option is available it will pay off handsomely.  The other pertains to the considerable computational difficulty (in many cases) of evaluating the evidence integrals.  If the statistical spread in the evidence is large, then only relatively crude estimates of these integrals may suffice as there is no need for an accurate calculation of a quality which will have large scatter for other reasons.  For instance, the Laplace approximation may suffice, or the Bayesian information criterion.

Finally, various authors have made connections between the Bayes factor and the classical $p$-value of significance testing.  Most recently Johnson \cite{jo13} has shown that convincing Bayes factors imply much smaller $p$-values than are usually accepted in many areas of science.  This goes some way to explaining why so many results cannot be reproduced, a claim made by Ioannidis \cite{io} in a provocative paper.  If convincing Bayes factors have to be boosted by their substantial statistical spread to be robust for decision making, as has been suggested here, then acceptable $p$-values have to be even smaller than Johnson suggests.  This is yet one more reason  why   Ioannidis \cite{io} may reasonably claim ``most published research findings are false''.

\begin{theacknowledgments}
I thank John Peacock, of the University of Edinburgh, for many probing and useful discussions of Bayesian methods.
\end{theacknowledgments}




\bibliographystyle{aipproc}   



\end{document}